\documentclass[fleqn,10pt]{SelfArx} % Document font size and equations flushed left

\usepackage[english]{babel} % Specify a different language here - english by default

\usepackage{lipsum} % Required to insert dummy text. To be removed otherwise
\usepackage{afterpage}

 \definecolor{color1}{RGB}{0,0,110} % Color of the article title and sections
\definecolor{color2}{RGB}{0,20,80} % Color of the boxes behind the abstract and headings

\usepackage{hyperref} % Required for hyperlinks
\hypersetup{hidelinks,colorlinks,breaklinks=true,urlcolor=color2,citecolor=color1,linkcolor=color1,bookmarksopen=false,pdftitle={Title},pdfauthor={Author}}

\newcommand{\blue}[1]{{\color{blue}{#1}}}
\newcommand{\red}[1]{{\color{red}{#1}}}

\JournalInfo{Uploaded to arXiv.org} % Journal information
\Archive{} % Additional notes (e.g. copyright, DOI, review/research article)

\PaperTitle{The effect of the charge pattern on the applicability of a nanopore as a sensor} % Article title

\Authors{Eszter M\'adai\textsuperscript{1,2}, M\'onika Valisk\'o\textsuperscript{1},  Dezs\H{o} Boda\textsuperscript{1}*} % Authors
\affiliation{\textsuperscript{1}\textit{Department of Physical Chemistry, University of Pannonia,  P. O. Box 158, H-8201 Veszpr\'em, Hungary}} % Author affiliation
\affiliation{\textsuperscript{2}\textit{Department of Material- and Geo-Sciences, Technische Universit\"{a}t  Darmstadt, Petersenstr.\ 23, D-64287 Darmstadt, Germany}} % 
\affiliation{*\textbf{Corresponding author}: dezsoboda@gmail.com} % Corresponding author

 % Keywords - if you don't want any simply remove all the text between the curly brackets
\newcommand{\keywordname}{Keywords} % Defines the keywords heading name

\Abstract{We investigate a model nanopore sensor that is able to detect analyte ions that are present in the electrolyte solution in very small concentrations.
The nanopore selectively binds the analyte ions with which the local concentrations of the ions of the background electrolyte (KCl), and, thus, the ionic current flowing through the pore is changed. 
Analyte concentration can be determined from calibration curves.
In our previous study (M\'{a}dai et al.  J. Chem. Phys., 147(24):244702, 2017.), we proposed a symmetric model (surface charge is negative all along the pore). 
The mechanism of sensing was a competition between K$^{+}$ and positive analyte ions, so increasing analyte concentration decreased K$^{+}$ current.
Here we allow asymmetric charge patterns on the pore wall (positive/negative/neutral along the pore), thus, gaining an additional device function, rectification, resulting in a dual responsive device.
We find that a bipolar nanopore is an efficient geometry with Cl$^{-}$ ions being the main charge carriers.
The mechanism of sensing is that more positive analyte ions attract more Cl$^{-}$ ions into the pore thus increasing the current.
Also they make the pore less asymmetric and, thus, decrease rectification.
We use a hybrid computer simulation method, where a generalization of the grand canonical Monte Carlo method to non-equilibrium (Local Equilibrium Monte Carlo) is coupled to the Nernst-Planck equation with which the flux is computed.}

\begin{document}

\flushbottom % Makes all text pages the same height

\maketitle % Print the title and abstract box

% \tableofcontents % Print the contents section

\thispagestyle{empty} % Removes page numbering from the first page

%----------------------------------------------------------------------------------------
%	ARTICLE CONTENTS
%----------------------------------------------------------------------------------------

%#######################################################################
%#######################################################################

\section{Introduction}
\label{sec:intro}

There has been a growing interest in designing and fabricating nanodevices for the purpose of chemical and biochemical sensing on the basis of the peculiar properties of nanopores \cite{sexton_mbs_2007,gyurcsanyi_trac_2008,howorka_csr_2009,piruska_csr_2010,makra_ecc_2014,shi_ac_2016,lepoitevin_acis_2017}.
One of these properties of nanopores is that their radius is measurable to the characteristic screening (Debye) length of the electrolyte.
This makes them ideal building blocks of nanodevices, because overlap of the double layers inside the pore makes sensitive tuning of ion concentrations inside the pore possible.
The device function (also called transfer function) is usually based on a mechanism in which the measurable output, the ionic current, depends on controllable parameters such as voltage or electrolyte composition.

Another important feature of nanopores is that their walls can be manipulated via chemical treatment.
Various charge patterns can be established along the nanopore \cite{stein_prl_2004,Siwy_2004,singh_jap_2011,nasir_acsami_2014,Zhang_2015} with proper functional groups that protonate or deprotonate at a given pH.
Interaction of ions with surface charges is a primary effect driving the conductance properties of the nanopore.

Chemical methods can also be used to tether molecules to the pore wall that bind other molecules (ions) selectively.
If these bound ions play the role of an analyte, and their binding events change the ionic current flowing through the pore, the nanopore can be used as a sensor.

A lot depends on the nature and strength of the binding.
If binding is strong, the analyte ions stay bound for a time comparable to the time of the measurement.
In this case, detection of these ions is based on studying transient phenomena.
One example is the case when the analyte ion is so large that its binding results in a detectable signal, a measurable change in electrical current (resistive pulse sensing \cite{bayley_chem_rev_2000,gyurcsanyi_trac_2008,makra_ecc_2014}).
Another example is when the concentration of analyte molecules is related to the time necessary to block the current \cite{siwy_jacs_2005}.

If binding energy is weaker, the analyte molecules bind and unbind many times during the time of measurement.
This results in a reversible binding process and a large number of binding/unbinding events, large enough to form and ensemble and to use statistical methods for quantitative analysis.
In these cases, called ensemble methods, individual binding/unbinding events cannot be detected.
Their effect, however, can be measured by relating analyte concentration to ionic current.

Because we use Monte Carlo (MC) simulations to model the nanopore sensor and compute probabilities, we are obviously in the realm of ensemble methods.
In our previous study \cite{madai_jcp_2017}, we proposed a simple sensor model, where the wall of the cylindrical nanopore was negatively charged and selective binding sites (modeled with short-range square-well (SW) potentials) were placed along the pore.
The main charge carrier was a monovalent cation, K$^{+}$, for example.
The basis of sensing was a competition between analyte ions (denoted by X$^{+}$) and K$^{+}$ ions for space inside the pore.
Larger concentration ($c_{\mathrm{X}}$), and, thus, larger chemical potential, of X$^{+}$ ions in the bulk resulted in an advantage in this competition.
If X$^{+}$ ions replace K$^{+}$ ions inside the pore, the current carried by the K$^{+}$ ions decreases.
This $c_{\mathrm{X}}$--sensitive competition, and, thus, $c_{\mathrm{X}}$--sensitive current was the basis of the mechanism of sensing.

The device function was the $I/I_{0}$ ratio, namely, the ratio of the currents in the presence and in the absence of X$^{+}$ ions.
In that symmetric setup, $I/I_{0}$ was the only device function.

In this extended study, we examine various asymmetric nanopore designs thus gaining an additional device function to detect the effect of the X$^{+}$ ions thus gaining a dual responsive device.
This device function is rectification that appears when the nanopore is asymmetric in its structural features \cite{siwy_nim_2003,hou_advmat_2010,cervera_ea_2011,zhang_cc_2013,ali_acsami_2015,zhang_csr_2018}, such as its charge pattern (e.g., bipolar nanopores) or geometry (e.g., conical nanopores). 
Rectification is defined as the ratio of current magnitudes in the ON (forward biased) and OFF (reversed biased) states taken at the two opposite signs of a given voltage, $|I(U)/I(-U)|$, where $|I(U)|>|I(-U)|$.
In general, rectification depends on voltage.

This work was considerably inspired by the studies of Ensinger et al.\ \cite{Ali_AC_2018,Ali_Lang_2017,Ali_ACSnano_2012,liu_JACS_2015,ensinger_2018} who used conical PET nanopores with inherent rectification properties.
Binding metal ions with functionalized surfaces changed the surface charge pattern on the nanopores' wall thus changing their conduction properties including rectification.
In another work \cite{ali_chemcomm_2015}, a cylindrical nanopore was treated with immobilized DNA aptamer (LyzAp–NH$_{2}$) and made negatively charged. 
The pore has been made asymmetric by exposing it to lysozime (Lyz) protein only on one side. 
This protein selectively binds to the DNA aptamer thus creating a positive surface charge on one side resulting in a bipolar pore. 
Using KCl solution of varying concentrations, it was shown that this bipolar pore rectifies with a rectification degree that is sensitive to both Lyz and KCl concentration. 

The paper of Vlassiouk et al.\ \cite{vlassiouk_jacs_2009} was also an inspiration. 
In that work, a geometrically asymmetric rectifying pore functionalized with $\gamma$DPGA antibodies was considered. 
A pH-dependent charge asymmetry was superimposed working against the geometrical asymmetry. 
Inversion of rectification as a function of pH was revealed. 
When $\gamma$DFGA glutamic acids were added, the charge pattern changed and a pH-dependent modulation of charges had the same rectification effect as the geometrical asymmetry. 
This is a dual responsive channel that is modulated by both pH and binding of an analyte molecule. 

Nanopore-based sensors relying on selective binding of various ions to functionalized pore surfaces are abundant in the literature.
The molecule that binds the targeted ion depends on the chemical specificity of that ion. 
For example, different active molecules are used for Li$^{+}$ \cite{Ali_AC_2018}, Cs$^{+}$ \cite{Ali_Lang_2017}, Ca$^{2+}$/Mg$^{2+}$ \cite{Ali_ACSnano_2012}, K$^{+}$ \cite{liu_JACS_2015,wu_langmuir_2017}, Na$^{+}$ \cite{liu_JACS_2015},  F$^{-}$  \cite{nie_chemsci_2015}, or Zn$^{2+}$ \cite{tian_chemcomm_2010}.
Especially sensitive sensors can be fabricated by using the high specificity of enzymatic recognition mechanisms \cite{ali_analchem_2011,hou_materchemA_2014,perezmitta_nl_2018}.
The spectrum of molecules that can be detected with such devices is wide from inorganic ions \cite{Ali_AC_2018,Ali_Lang_2017,Ali_ACSnano_2012,liu_JACS_2015,wu_langmuir_2017,nie_chemsci_2015,tian_chemcomm_2010}, through amino acids \cite{ali_chemcomm_2015,vlassiouk_jacs_2009}, to sugars \cite{sun_chemcomm_2012}.
The correct detection of small amounts of such molecules and ions has practical significance in manufacturing engineering of high hydrogen and energy content alternative fuel components from bio resources \cite{hancsok_2011}.

In this modeling study, specificity of the binding potential is taken into account simply by defining the SW potential to act only on the X$^{+}$ ions.
In the continuum water framework applied in this study, model parameters include ion charges, ion radii, pore radius, pore length, pore charges, ion concentrations, and voltage (here, we change only pore charges, voltage, and X$^{+}$ concentration).
This modeling level including continuum solvent and hard sphere ions is especially appropriate for the nanopore sensor studied here for the following reasons.
(1) Proper description of the competition between ions requires accurate computation of ionic correlations including their finite sizes. 
Computer simulation of ions with finite diameters in a non-equilibrium system, therefore, is necessary.
Simulation is also useful if we want to consider complex geometries and interparticle potentials. 
(2) We need to simulate ions of very low (micromolar) concentrations. 
This requires implicit solvent. 
While Brownian dynamics (BD) would be an obvious choice \cite{berti_jctc_2014}, simulation of trace concentrations requires a grand canonical Monte Carlo (GCMC) scheme with ion insertion/deletions.

We have proposed a solution for this challenge \cite{boda_jctc_2012}: the Local Equilibrium Monte Carlo (LEMC) method coupled to the Nernst-Planck (NP) transport equation with which the ionic flux is computed.
This hybrid method, called NP+LEMC, was applied for various problems in the last couple of years.
These problems include particle transport through model membranes \cite{boda_jctc_2012,hato_jcp_2012}, ion channels \cite{boda_jml_2014,boda_arcc_2014,hato_cmp_2016}, and nanopores \cite{hato_pccp_2017,matejczyk_jcp_2017,madai_jcp_2017,fertig_mp_2018,madai_pccp_2018}.
With this computational technique, we are able to simulate the effect of even a very low concentration of X$^{+}$ ions on the current through the nanopore \cite{madai_jcp_2017}.

Our systematic study presents different combinations of positive, negative, and zero surface charges in the left and right regions of a nanopore.
We present results for uniformly charged, unipolar, and bipolar nanopores.
We discuss advantages and disadvantages of the various combinations and show that a bipolar pore proves to be the most appropriate design next to our original design (uniformly charged negative pore \cite{madai_jcp_2017}).
We show that fundamentally different mechanisms of sensing work in the two cases.

\section{Model and method}
\label{sec:model}

\subsection{Nanopore model}
\label{subsec:pore}

\begin{figure}[t]
	\begin{center}
		\includegraphics*[width=0.5\textwidth]{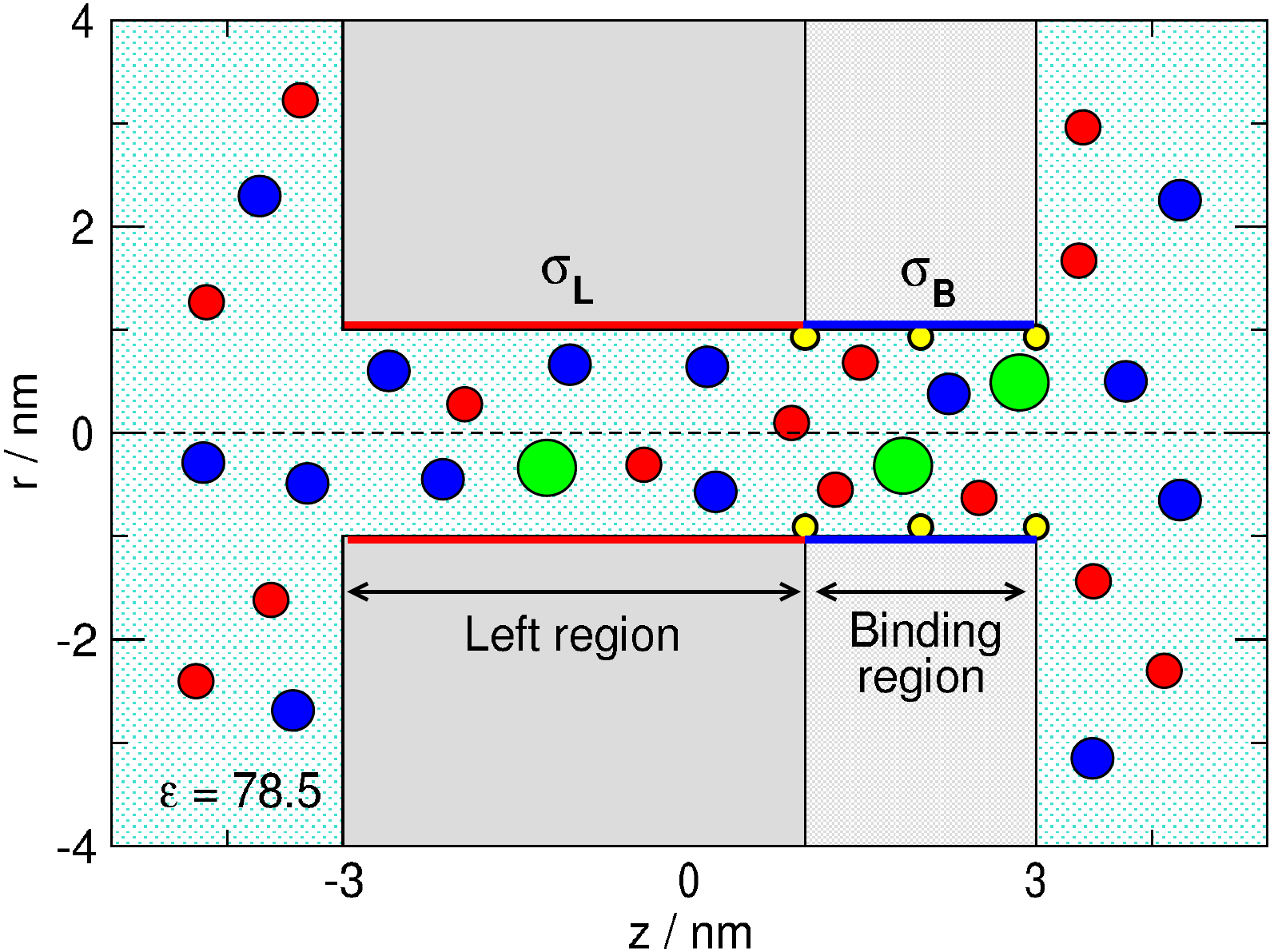}
		\caption{\small Schematics of the cylindrical pore that is divided into two regions. 
		The left region of length $H_{\mathrm{L}}=4$ nm carries $\sigma_{\mathrm{L}}$ surface charge that determines the main charge carrier of the pore.
		The right region is the ``binding region'', where the binding sites are located (indicated by small yellow spheres on the wall).
		The length of this region is $H_{\mathrm{B}}=2$ nm and can carry $\sigma_{\mathrm{B}}$ surface charge.
		The radius of the pore $R_{\mathrm{pore}}=1$ nm in this study.
		The green sphere is the analyte ion, X$^{+}$, that is bound to the binding site if it overlaps with the yellow sphere. 
		The blue and red spheres are the cations (K$^{+}$) and anions (Cl$^{-}$) of the electrolyte, respectively.
		}
		\label{fig1}
	\end{center}
\end{figure}

The nanopore is a cylindrical pore of radius $R_{\mathrm{pore}}=1$ nm penetrating a membrane of width 6 nm (Fig.\ \ref{fig1}).
The walls of the pore and the membrane are hard, namely, overlap of ions with these walls is forbidden.
The pore is divided into two regions along the $z$-axis  ($z$ is the coordinate along the main axis of the pore, perpendicular to the membrane).
The left region carrying $\sigma_{\mathrm{L}}$ surface charge has a length $H_{\mathrm{L}}=4$ nm. 
This is the region whose surface charge determines the main charge--carrier ionic species.

The right region, called binding region, contains the binding sites  and carries $\sigma_{\mathrm{B}}$ surface charge.
Its length, $H_{\mathrm{B}}=2$ nm, is half of the left region in this study.

The surface charge is represented by fractional point charges that are situated on a rectangular grid, where a surface element is approximately a square of size $0.2 \times 0.2$ nm$^{2}$.
The magnitude of the point charges is established so that the surface charge density corresponds to the prescribed values, $\sigma_{\mathrm{L}}$ and $\sigma_{\mathrm{B}}$.

We investigated various combinations depending whether the values of surface charges take the value of 1 $e$\,nm$^{-2}$ (denoted by \red{\bf p} and red color in the figures), 0 $e$\,nm$^{-2}$ (denoted by \textbf{0} and black color in the figures), or -1 $e$\,nm$^{-2}$ (denoted by \blue{\bf n} and blue color in the figures).
The value of the $\sigma_{\mathrm{L}}$ charge is either $-1$ or $1$ $e$\,nm$^{-2}$, while the value of the $\sigma_{\mathrm{B}}$ charge can take either $-1$, $0$, or $1$ $e$\,nm$^{-2}$. 
That makes six combinations denoted by \blue{\bf nn}, \blue{\bf n}\textbf{0}, \blue{\bf n}\red{\bf p}, \red{\bf pp}, \red{\bf p}\textbf{0}, and \red{\bf p}\blue{\bf n}.
In the main text, we will use these notations without colors from now on.

\subsection{Interparticle potentials}

The electrolyte model is the ``primitive'' model  that represents ions as charged hard spheres of charges, $q_{i}$, and radii, $R_{i}$:
\begin{equation}
u_{ij}(r) =
\left\lbrace 
\begin{array}{ll}
\infty & \quad \mathrm{for} \quad r<R_{i}+R_{j} \\
 \dfrac{1}{4\pi\epsilon_{0}\epsilon} \dfrac{q_{i}q_{j}}{r} & \quad \mathrm{for} \quad r \geq R_{i}+R_{j}\\
\end{array}
\right. 
\label{eq:uij}
\end{equation} 
where $\epsilon_{0}$ is the permittivity of vacuum, $\epsilon$ is the dielectric constant of the electrolyte, and $r$ is the distance between two ions.
Subscript $i$ can take values $i=+$, $-$, and X, where these symbols refer to the cation of the electolyte (K$^{+}$), the anion of the electrolyte (Cl$^{-}$), and the analyte ion (X$^{+}$), respectively.
The ionic radii are $R_{+}=0.133$ nm, $R_{-}=0.181$ nm (Pauling radii), and $R_{\mathrm{X}}=0.3$ nm, in this study.
The concentration of the KCl background electrolyte is fixed at $0.01$ M.

The solvent is represented as a continuum background characterized by two response functions. 
One is the dielectric constant, $\epsilon=78.5$, that describes the screening effect of the water molecules.
The other is a diffusion coefficient function, $D_{i}(z)$, that describes the ability of water molecules to affect the diffusion of ions.
This function is space dependent in our case; it is a piecewise constant function that is different inside the pore ($D_{i}^{\mathrm{pore}}$)
and in the bulk ($D_{i}^{\mathrm{bulk}}$).
The bulk value is experimental ($D_{\mathrm{K}^{+}}=1.849\times10^{-9}$ m$^{2}$s$^{-1}$ and $D_{\mathrm{Cl}^{-}}=D_{\mathrm{X}^{+}}=2.032\times10^{-9}$  m$^{2}$s$^{-1}$), while $D_{i}^{\mathrm{pore}}$ just scales the current without influencing the $I/I_{0}$ ratio.
Following our previous studies \cite{matejczyk_jcp_2017,madai_jcp_2017,fertig_mp_2018,madai_pccp_2018} here we set the relation $D_{i}^{\mathrm{pore}}=0.1\,D_{i}^{\mathrm{bulk}}$.
In other studies, where reference was available, we fitted the $D_{i}^{\mathrm{pore}}$ value to experimental \cite{boda_arcc_2014,fertig_mp_2018} or molecular dynamics \cite{hato_pccp_2017} data.

We placed the binding sites on the pore wall in $3$ rings placed at $z=1$, $2$, and $3$ nm.
Each ring contains 4 binding sites \cite{madai_jcp_2017}.
The binding potential between a site and an analyte ion is the square-well (SW) potential:
\begin{equation}
u_{\mathrm{SW}}(r) =
\left\lbrace 
\begin{array}{ll}
0 & \quad \mathrm{for} \quad r-R_{\mathrm{X}}>d_{\mathrm{SW}}\\
 -\epsilon_{\mathrm{SW}} & \quad \mathrm{for} \quad r-R_{\mathrm{X}}<d_{\mathrm{SW}} ,
\end{array}
\right. 
\label{eq:squarewell}
\end{equation}
where $r$ is the distance of the site and the ion center.
This short-range potential attracts X$^+$ with $-\epsilon_{\mathrm{SW}}=-10\,kT$ energy once the closest point of the X$^+$ ion's surface is closer to the site than the distance parameter $d_{\mathrm{SW}}=0.2$ nm.
This model takes into account that the active site of the X$^+$ ion is usually on its surface while keeping the spherical symmetry of the ion (it neglects the possible orientation dependence of binding).
The SW potential acts only on the X$^{+}$ ions in the simulations. 

\subsection{NP+LEMC}
\label{sec:method}

In the NP+LEMC  technique \cite{boda_jctc_2012} the NP equation is used to compute the ionic flux for ion species $i$
\begin{equation}
 \mathbf{j}_{i}(\mathbf{r})=-\dfrac{1}{kT}D_{i}(\mathbf{r})c_{i}(\mathbf{r}) \nabla \mu_{i}(\mathbf{r}) ,
\label{eq:np}
\end{equation} 
where $T=298.15$ K is temperature, $k$ is Boltzmann's constant,  $c_{i}(\mathbf{r})$ is the concentration profile, $\mu_{i}(\mathbf{r})$ is the electrochemical potential profile, and $\mathbf{j}_{i}(\mathbf{r})$ is the particle flux density.

Solution of the NP equation requires a relation between $c_{i}(\mathbf{r})$ and $\mu_{i}(\mathbf{r})$.
Here we use the Local Equilibrium Monte Carlo (LEMC) simulation method that is an adaptation of the GCMC technique to a non-equilibrium situation.
Due to its grand canonical nature, LEMC can handle small concentrations easily.
 
We divide the computation domain of the NP system into volume elements,  $\mathcal{D}^{\alpha}$, and use different $\mu_{i}^{\alpha}$ values in each volume element.
Insertions/deletions of ions are attempted into/from these volume elements with equal probability.
These trials are accepted or refused on the basis of the Metropolis algorithm \cite{metropolis}.
The acceptance probability contains the local electrochemical potential, $\mu_{i}^{\alpha}$, and the energy change.
The energy includes every interaction from the whole simulation cell, not only from subvolume $\mathcal{D}^{\alpha}$.
The result of the LEMC simulation in an iteration is the concentration in every volume element, $c_{i}^{\alpha}$.

The whole system is solved in an iterative way by adjusting the electrochemical potential profile ($\mu_{i}^{\alpha}$) in each iteration until conservation of mass ($\nabla \cdot \mathbf{j}_{i}(\mathbf{r})=0$) is satisfied.
Transport described by the NP and the continuity equations and statistical mechanics handled with LEMC form a self-consistent system.
The advantage over the widely used Poisson-Nernst-Planck theory is that the statistical mechanical component is an accurate particle simulation method instead of  an approximate mean--field theory (Poisson-Boltzmann) \cite{matejczyk_jcp_2017,madai_pccp_2018}. 
Details are found in earlier papers  \cite{boda_jctc_2012,boda_jml_2014}.

The computational domain is a closed system (no periodic boundary conditions are applied).
We apply boundary conditions for concentration (chemical potential) and electrical potential on the boundary of our cell that is a cylinder in this study. 
Different constant values are applied on the two sides of the membrane at the boundaries of the two half-cylinders as described in our previous studies \cite{boda_jctc_2012,boda_jml_2014}.
Since the boundary conditions are fixed, the driving force of the transport is maintained, so the transport is steady state. 

\section{Results}
\label{sec:res}

In our previous study \cite{madai_jcp_2017}, we had a negatively charged pore (nn) with the binding sites placed in the middle of the pore, therefore, the pore was symmetric in structure.
Consequently, we had only one device function $I/I_{0}$, where $I$ and $I_{0}$ are the currents in the presence and absence of X$^{+}$ ions, respectively.

Here, the pore is asymmetric because of the asymmetric placement of the binding sites in only the right region.
This choice reflects the experimental setup where binding sites are placed at the tip of conical nanopores \cite{vlassiouk_jacs_2009,ali_chemcomm_2015}.
Asymmetric charge patterns can superimpose on this asymmetry.
Therefore, rectification as an additional device function appears.
Rectification is defined as $|I_{\mathrm{ON}}|/|I_{\mathrm{OFF}}|$, where $I_{\mathrm{ON}}$ and $I_{\mathrm{OFF}}$ are the currents in the open and closed states of the pore, respectively.
The ON and OFF states are defined by the relation $|I_{\mathrm{ON}}|>|I_{\mathrm{OFF}}|$.

We provide a systematic study for various combinations of the $\sigma_{\mathrm{L}}$ and $\sigma_{\mathrm{B}}$ surface charges and the effects of these surface charge patterns on the efficiency of a nanopore-based sensor, where binding sites that selectively bind the analyte ions (X$^{+}$) are placed in the right ($\sigma_{\mathrm{B}}$) region.
The left region is either \textbf{n} or \textbf{p}, while the binding region can be \textbf{n}, \textbf{0}, or \textbf{p}.
These six combinations denoted by \textbf{nn}, \textbf{n0}, \textbf{np}, \textbf{pp}, \textbf{p0}, and \textbf{pn} have been studied by performing simulations for varying voltages and $X^{+}$ concentrations.

\begin{figure*}[ht!]
	\begin{center}
		\includegraphics*[width=0.76\textwidth]{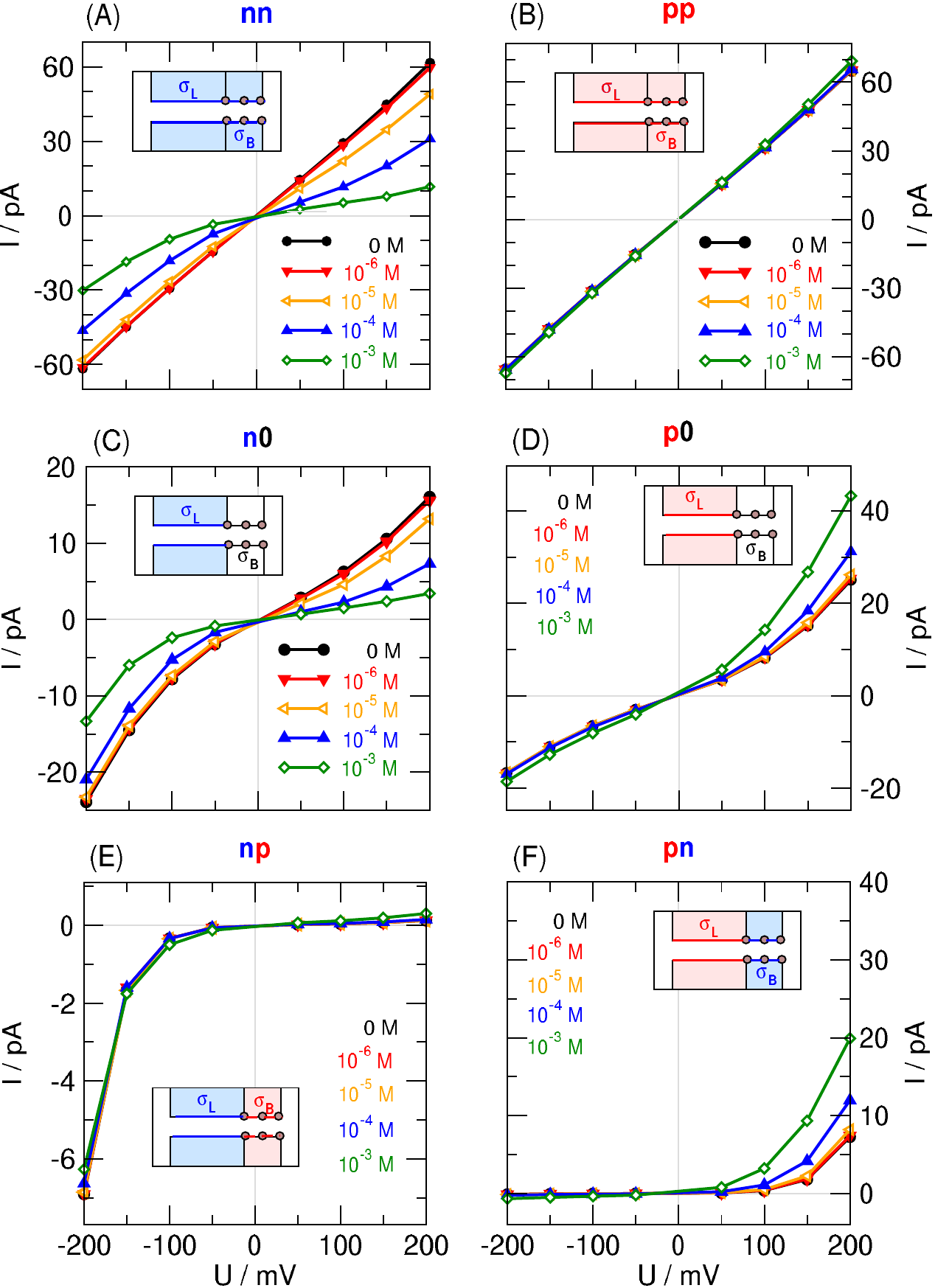}
		\caption{\small Current-voltage curves ($I$ vs.\ $U$) for the different charge patterns. 
		Left panels (A, C, and E) refer to charge patterns, where the left region is negative, $\sigma_{\mathrm{L}}=-1$ $e$/m$^{2}$, while right panels (B, D, and F) refer to charge pattern, where the left region is positive, $\sigma_{\mathrm{L}}=1$ $e$/m$^{2}$.
		Top panels (A and B) refer to uniformly charged pores, \textbf{nn} and \textbf{pp}.
		Middle panels (C and D) refer to unipolar pores, \textbf{n0} and \textbf{p0}.
		Bottom panels (E and F) refer to bipolar pores, \textbf{np} and \textbf{pn}.
		For explanation of notations, see the main text.
		The values of $\sigma_{\mathrm{B}}$ and $\sigma_{\mathrm{L}}$ take the values -1, 0, or 1 $e$/nm$^{2}$.
		Every panel shows results for different X$^{+}$ concentrations from 0 to 10$^{-3}$ M. 
		}
		\label{fig2}
	\end{center}
\end{figure*}

Figure \ref{fig2} shows the current vs.\ voltage ($I-U$) curves for all charge patterns and X$^{+}$ concentrations from 0 to $10^{-3}$ M.
These are the raw data obtained from the simulations and also the primary results produced by experiments.
The diverse behavior of the $I-U$ curves as a function of the various charge patterns is apparent.
Also, the effect of the varying concentration of the X$^{+}$ ions can be deduced from these plots.
The color code described in subsection \ref{subsec:pore} is shown by the sketches of the various geometries in Fig.\ \ref{fig2}.

% The $\sigma_{\mathrm{L}}$ charge of the wider left region determines the main charge carrier of the nanopore.
% The $\sigma$

% The ON and OFF states are measured at $\pm 200$ mV voltage.
% Which sign of the voltage produced the ON state depended on the charge distribution.

% \afterpage{\clearpage}

\begin{figure*}[ht!]
	\begin{center}
		\includegraphics*[width=0.758\textwidth]{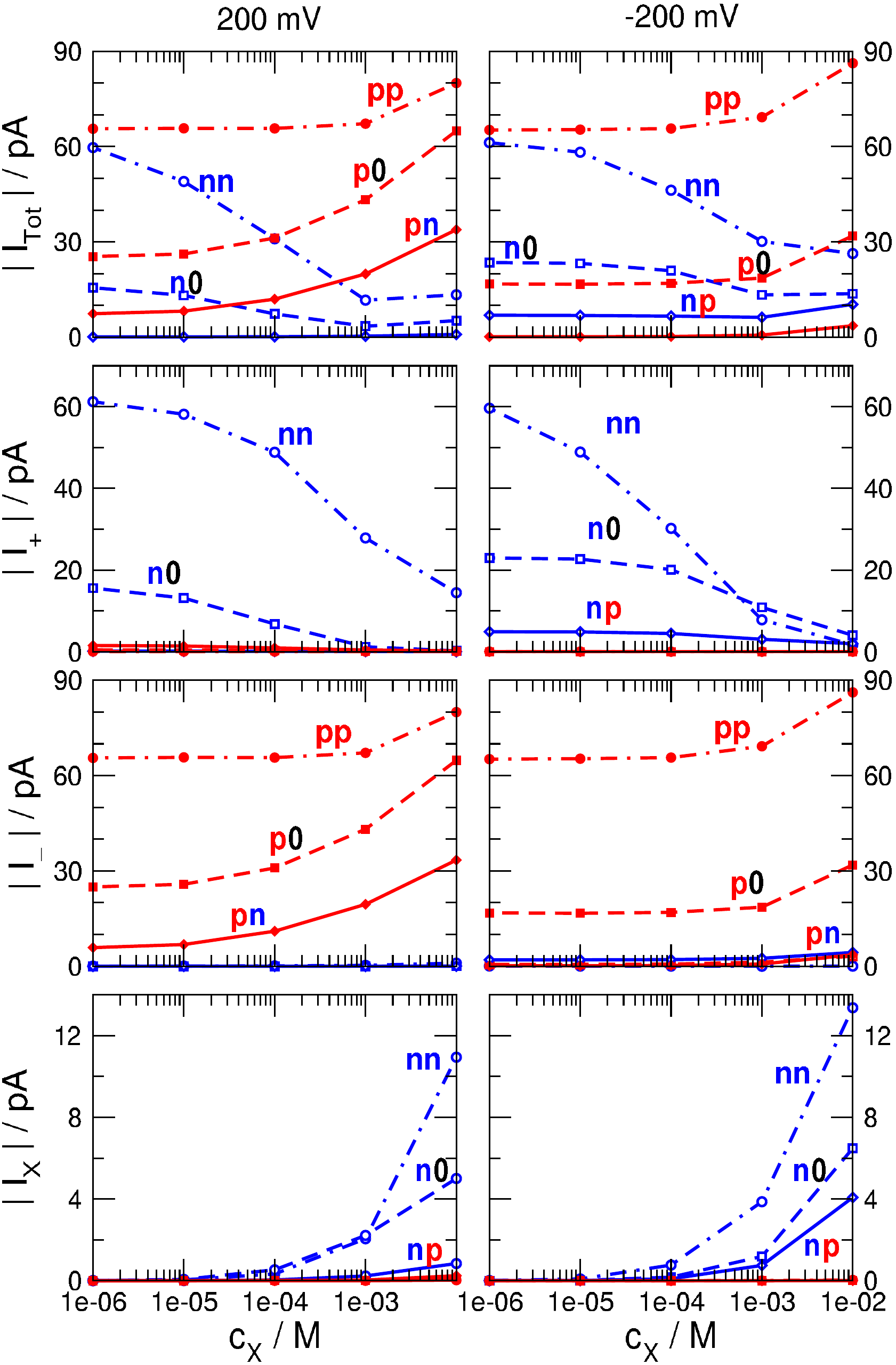}
		\caption{\small
		Absolute values of currents for two selected voltages (200 mV and -200 mV in left and right panels, respectively) as functions of the X$^{+}$ concentration.
		The top panels show the total currents, while panels below it from top to bottom show currents carried by the cations, anions, and X$^{+}$.
		The \textbf{nn} and \textbf{pp} charge patterns are shown by dot-dashed lines, the \textbf{n0} and \textbf{p0} charge patterns by dashed lines, while \textbf{np} and \textbf{pn} charge patterns by solid lines.
		The \textbf{nn}, \textbf{n0}, and \textbf{np} charge patterns are shown in blue with open symbols, while the \textbf{pp}, \textbf{p0}, and \textbf{pn} charge patterns in red with filled symbols.
		The figure indicates that for the \textbf{n0} and \textbf{np} charge patterns -200 mV is the ON state, while for the \textbf{p0} and \textbf{pn} charge patterns 200 mV is the ON state.
		}
		\label{fig3}
	\end{center}
\end{figure*}

% \afterpage{\clearpage}

\begin{figure}[t!]
	\begin{center}
		\includegraphics*[width=0.45\textwidth]{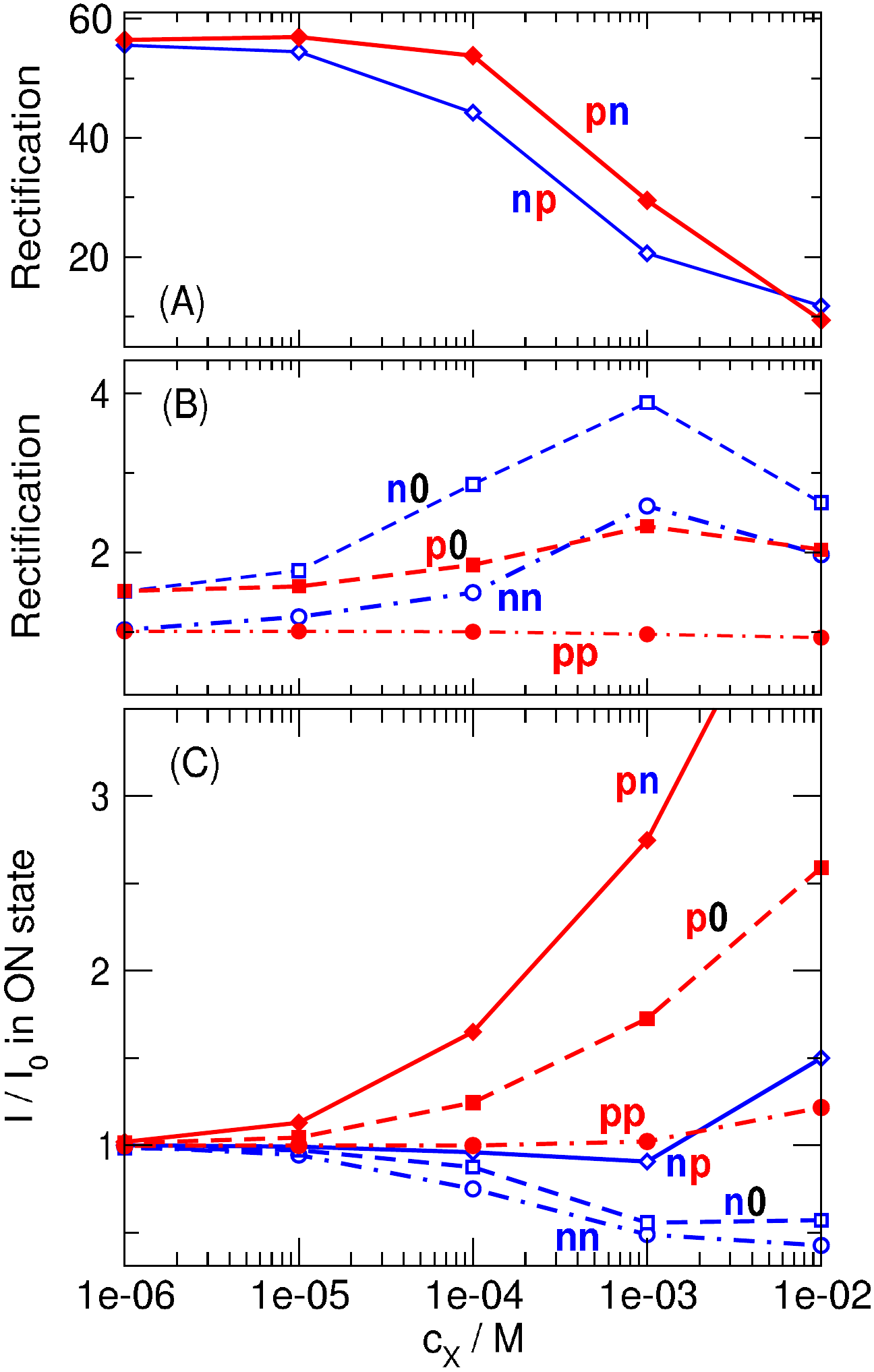}
		\caption{\small 
		The dependence of the device functions on c$_{\mathrm{X}}$ for the various charge patterns.
		Panels A and B show the rectification defined as $|I_{\mathrm{ON}}/I_{\mathrm{OFF}}|$ (for the \textbf{n0} and \textbf{np} charge patterns -200 mV is the ON state, while for the \textbf{p0} and \textbf{pn} charge patterns 200 mV is the ON state).
		Panel A shows the large rectifications characteristic of the bipolar pores (np and \textbf{pn}).
		Panel B shows the rectification for the other geometries.
		Panel C shows the ON-state $I/I_{0}$ ratio with $I_{0}$ being the current at $c_{\mathrm{X}}=0$ M.
% 		The ON state is either 200 mV or -200 mV depending which one results in larger currents.
		}
		\label{fig4}
	\end{center}
\end{figure}

Although the shape of the $I-U$ curve can also be considered as a device function, we are interested in a more quantitative analysis.
Therefore, we choose two representative voltages ($\pm 200$ mV) and show detailed results for these cases. 
Figure \ref{fig3} shows the magnitudes of currents as functions of $c_{\mathrm{X}}$ for voltages $\pm 200$ mV.
Total currents (top row) and currents carried by the individual ionic species (K$^{+}$, Cl$^{-}$, and X$^{+}$ from top to bottom) are shown.
Figure \ref{fig4} shows the device functions plotted against analyte concentration ($c_{\mathrm{X}}$).
The $I/I_{0}$ ratios are shown for the ON states, because currents are larger in those cases.
The voltage -200 mV is the ON state for charge patterns \textbf{nn}, \textbf{n0}, and \textbf{np}, while 200 mV is the ON state for charge patterns \textbf{pp}, \textbf{p0}, and \textbf{pn}.
In Fig.\ \ref{fig4}, one of our interest is the sensitivity of the device that we define as the response (either $I/I_{0}$ or rectification) given to a given certain degree of input signal ($c_{\mathrm{X}}$).
We can also define sensitivity as the slope of the device function vs. $c_{\mathrm{X}}$ plot.

The \textbf{nn} case practically corresponds to the geometry of our previous work \cite{madai_jcp_2017} except for the fact that the binding sites are placed asymmetrically here (Fig.\ \ref{fig2}A).
Accordingly, current decreases with increasing $c_{\mathrm{X}}$ as in our previous work ($I_{\mathrm{tot}}$ and $I_{+}$ in Fig.\ \ref{fig3}).
This type of the sensor works on the basis of a competition between the two cationic species, K$^{+}$ and X$^{+}$.
Therefore, while the current of K$^{+}$ decreases with increasing $c_{\mathrm{X}}$, an increase of the X$^{+}$ current can be observed especially when $c_{\mathrm{X}}$ is large enough.
The increase of X$^{+}$ current counteracts the decrease of the total (measurable) current (Fig.\ \ref{fig3}), therefore, leaking X$^{+}$ current makes this setup a less efficient sensor on the basis of the $I/I_{0}$ ratio.
Note that the total current shows a better sensitivity to $c_{\mathrm{X}}$ in the OFF state (200 mV). 
As expected, rectification is small, so no additional device function appears in this case (Fig.\ \ref{fig4}B).

The sign of the charge of the X$^{+}$ ion makes the difference between the charge patterns regarding the sign of either $\sigma_{\mathrm{L}}$ or $\sigma_{\mathrm{B}}$.
As a test, we performed simulations for the \textbf{pp} case (Fig.\ \ref{fig2}B), where the pore is positively charged. 
Since the X$^{+}$ ions are repulsed by the surface charges, they do not enter the pore, so they do not influence the $I-U$ curves (Fig.\ \ref{fig2}B).
Consequently, this nanopore cannot be used as a sensor (Figs.\ \ref{fig4}B and C).

When the surface charges on the right hand side are removed (e.g., $\sigma_{\mathrm{B}}=0$), we obtain unipolar nanopores, \textbf{n0} and \textbf{p0}.
These are rectifying pores, but their rectifications are so small that they do not provide rectification as an additional device function (Fig.\ \ref{fig4}B), at least, for the geometry studied here..
As far as the $I/I_{0}$ ratio is concerned, the \textbf{n0} and \textbf{p0} geometries behave differently.

In the \textbf{n0} geometry (Fig.\ \ref{fig2}C), the X$^{+}$ ions are not attracted by the $\sigma_{\mathrm{B}}=0$ surface charge, so they are more reluctant to enter the pore and bind to the binding sites.
Therefore, this sensor is less sensitive than the \textbf{nn} sensor, especially at small X$^{+}$ concentrations that are our main interest (Fig.\ \ref{fig4}C). 
Because rectification is too small to be usable as a device function, this geometry is not an advance compared to the \textbf{nn} geometry.

In the \textbf{p0} geometry (Fig.\ \ref{fig2}D), rectification does not depend on the analyte concentration (Fig.\ \ref{fig4}B), but the $I/I_{0}$ ratio increases as $c_{\mathrm{X}}$ increases (Fig.\ \ref{fig4}C).
The explanation is that this pore is better in accepting and binding the X$^{+}$ ions compared to the \textbf{pp} pore, because the neutral $\sigma_{\mathrm{B}}$ region now does not repel the X$^{+}$ ions.
As X$^{+}$ ions bind to the binding sites, they attract more Cl$^{-}$ ions that are the main charge carriers. 
This geometry has the advantage of excluding the X$^{+}$ ions from the $\sigma_{\mathrm{L}}$ region and preventing their leakage--current (Fig.\ \ref{fig3}).
The increase of $c_{\mathrm{X}}$, therefore, influences the total current only via influencing the current of Cl$^{-}$ ions. 
This geometry, therefore, makes a good sensor regarding the $I/I_{0}$ ratio but without the additional benefit from pore asymmetry and rectification (Fig.\ \ref{fig4}B-C).

As expected, rectification appears as an additional device function when we create bipolar pores: \textbf{np} or \textbf{pn}.
The \textbf{np} geometry (Fig.\ \ref{fig2}E), as expected, is not a help, because X$^{+}$ ions are repulsed by the positive $\sigma_{\mathrm{B}}$ charge, so the pore is not sensitive to the presence of X$^{+}$ ions.
Although the rectification shows a reasonable $c_{\mathrm{X}}$ dependence (Fig.\ \ref{fig4}A), the magnitudes of currents are small (much smaller than in the \textbf{nn} case, Fig.\ \ref{fig2}E vs.\ A) and there is considerable X$^{+}$ leakage (Fig.\ \ref{fig3}).
Small currents are not advantageous for actual devices due to possible problems regarding signal/noise relation.

\begin{figure*}[t]
	\begin{center}
		\includegraphics*[width=0.6\textwidth]{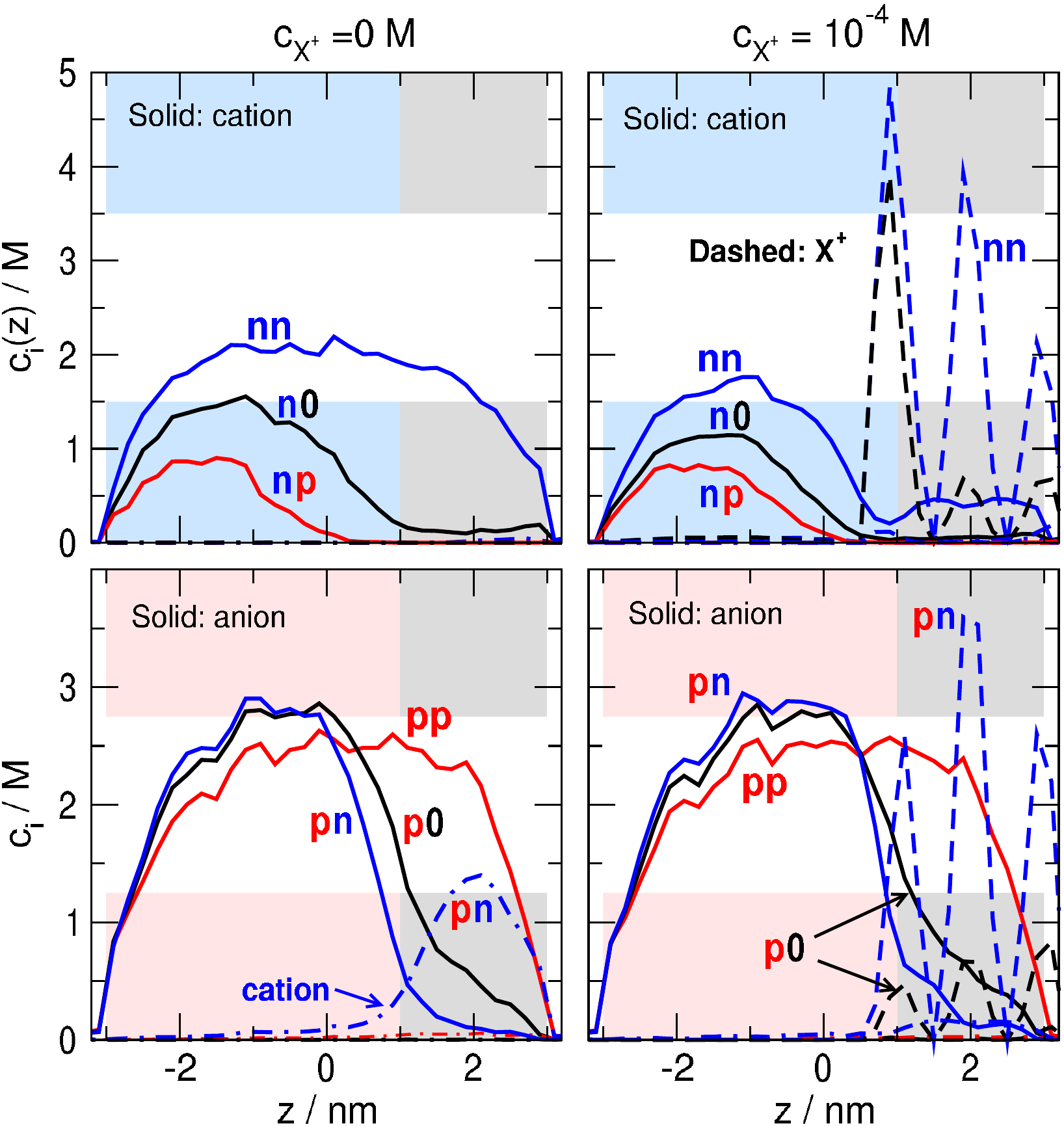}
		\caption{\small 
		Concentration profiles for $c_{\mathrm{X}}=0$ M (left panels) and $c_{\mathrm{X}}=10^{-4}$ M (right panels) for the ON states.
		Profiles for X$^{+}$ are shown with dashed lines.
		Solid lines show the profiles for the main charge carriers, while the dot-dashed lines show the profiles for the other ion.
		Top panels show the results for the \textbf{nn}, \textbf{n0}, and \textbf{np} geometries with solid lines being the cation profiles.
		Bottom panels show the results for the \textbf{pp}, \textbf{p0}, and \textbf{pn} geometries with solid lines being the anion profiles.
		The colors are determined by the $\sigma_{\mathrm{B}}$ charge (blue for n, black for 0, and red for p).
		}
		\label{fig5}
	\end{center}
\end{figure*}

The \textbf{pn} geometry (Fig.\ \ref{fig2}F), on the other hand exhibits all the useful behavior that we expect from our device.
First, the negative $ \sigma_{\mathrm{B}}$ surface charge attracts the X$^{+}$ ions into the $\sigma_{\mathrm{B}}$ region increasing the probability of their presence and binding to the binding sites (Fig.\ \ref{fig3}).
This increases the sensitivity of the device to the concentration of X$^{+}$ ions.
Second, there is a large rectification that is strongly dependent on $c_{\mathrm{X}}$. 
Rectification decreases as $c_{\mathrm{X}}$ increases because the accumulation of the positive X$^{+}$ ions in the right region impairs the \textbf{pn} charge asymmetry of the nanopore (Fig.\ \ref{fig4}A).
Third, the total current (and, consequently, the $I/I_{0}$ ratio) has a steady increase as a function of $c_{\mathrm{X}}$ in the ON state (Fig.\ \ref{fig3} and \ref{fig4}C).

In the absence of X$^{+}$, this is mainly anion current, because the $\sigma_{\mathrm{L}}$ region is longer than the $\sigma_{\mathrm{B}}$ region.
The ratio of the anion vs.\ cation currents can be adjusted with the lengths of the two regions, $H_{\mathrm{L}}$ and $H_{\mathrm{B}}$.
As the concentration of X$^{+}$ is increased, cation currents decrease because the X$^{+}$ ions outcompete the K$^{+}$ ions in the right region.
In the meantime, the presence of X$^{+}$ ions in the right regions increase the concentration of the Cl$^{-}$ ions in the right region thus increasing anion current.

All these effects are well visible in Fig.\ \ref{fig5} where the ON-state concentration profiles are plotted for two analyte concentrations ($c_{\mathrm{X}}=0$ and $10^{-4}$ M).
The top panels show the results for the cases, when the cations are the main charge carriers ($\sigma_{\mathrm{L}}=-1$ $e$/nm$^{2}$).
As discussed above, only the \textbf{nn} geometry behaves as a decent sensor for this case (considered in our previous work  \cite{madai_jcp_2017}).
The X$^{+}$ ions accumulate on the right hand side at $c_{\mathrm{X}}=10^{-4}$ M while decreasing the K$^{+}$ concentrations there.
This results in a decrease of the total current.

The bottom panels show the results for the cases, when the anions are the main charge carriers ($\sigma_{\mathrm{L}}=1$ $e$/nm$^{2}$).
In this case, the \textbf{pn} geometry shows considerable binding of X$^{+}$ ions for $c_{\mathrm{X}}=10^{-4}$ M.
They squeeze the cations out and attract anions in.

This latter effect is the basis of sensor function.
While in the \textbf{nn} case a $c_{\mathrm{X}}$-dependent competition between K$^{+}$ and X$^{+}$ ions was the basis of sensing the X$^{+}$ ions, here an attraction between X$^{+}$ ions and the charge--carrier Cl$^{-}$ ions is the basis of the $c_{\mathrm{X}}$--sensitive effect.
This is primarily an electrostatic effect.
The X$^{+}$ ions are attracted by the negative  $\sigma_{\mathrm{B}}$ surface charge, but only to the right region.
The X$^{+}$ ions tune both the Cl$^{-}$ concentration in the right region and the asymmetry of the pore.
This way, both the $I/I_{0}$ ratio and rectification can be used as a device function. 

\section{Summary}

Inspired by experimental works in the field of nanosensing \cite{Ali_AC_2018,Ali_Lang_2017,Ali_ACSnano_2012,liu_JACS_2015,ensinger_2018,ali_chemcomm_2015, vlassiouk_jacs_2009}, we continued our line of investigation \cite{madai_jcp_2017} and constructed asymmetric nanopores with various charge patterns. 
Working with asymmetric nanopores gives rise to an additional device function, rectification, that makes it possible to determine analyte concentration on the basis of two variables instead of just one.

We found that the \textbf{pn} geometry (making a bipolar nanopore) is an appropriate setup for several reasons.
The analyte ions are attracted electrostatically (in addition to the square-well short-range attraction), so the pore is sensitive to the presence of the X$^{+}$ ions.
The X$^{+}$ ions exert their effect on the main charge carriers (anions) by attracting them.
Rectification is also a proper device function, because  the pore becomes less asymmetric as more X$^{+}$ ions are bound.
Rectification, therefore, decreases with increasing X$^{+}$ concentration.

In this study we focused on the effect of charge pattern and fixed most of the parameters of our model sensor.
In a subsequent study, we present results of extensive simulations for the \textbf{pn} geometry by changing experimentally controllable parameters (pore radius, KCl concentration, charge of analyte ion, and pH) in order to explore the capabilities of this seemingly promising geometry as a sensor.

\section*{Acknowledgments}
 
We gratefully acknowledge  the financial support of the National Research, Development and Innovation Office -- NKFIH K124353. 
Present article was published in the frame of the project GINOP-2.3.2-15-2016-00053.

% \bibliography{nanopore,book,own}

\begin{thebibliography}{10}

\bibitem{sexton_mbs_2007}
L.~T. Sexton, L.~P. Horne, and C.~R. Martin.
\newblock Developing synthetic conical nanopores for biosensing applications.
\newblock {\em Mol. BioSyst.}, 3(10):667--685, 2007.

\bibitem{gyurcsanyi_trac_2008}
R.~E. Gyurcs{\'{a}}nyi.
\newblock Chemically-modified nanopores for sensing.
\newblock {\em {TrAC}-Trend. Anal. Chem.}, 27(7):627--639, 2008.

\bibitem{howorka_csr_2009}
S.~Howorka and Z.~Siwy.
\newblock Nanopore analytics: sensing of single molecules.
\newblock {\em Chem. Soc. Rev.}, 38(8):2360--2384, 2009.

\bibitem{piruska_csr_2010}
A.~Piruska, M.~Gong, and J.~V. Sweedler.
\newblock Nanofluidics in chemical analysis.
\newblock {\em Chem. Soc. Rev.}, 39:1060--1072, 2010.

\bibitem{makra_ecc_2014}
I.~Makra and R.~E. Gyurcs{\'{a}}nyi.
\newblock Electrochemical sensing with nanopores: A mini review.
\newblock {\em Electrochem. Commun.}, 43:55--59, 2014.

\bibitem{shi_ac_2016}
W.~Shi, A.~K. Friedman, and L.~A. Baker.
\newblock Nanopore sensing.
\newblock {\em Anal. Chem.}, 89(1):157--188, 2016.

\bibitem{lepoitevin_acis_2017}
M.~Lepoitevin, T.~Ma, M.~Bechelany, J.-M. Janot, and S.~Balme.
\newblock Functionalization of single solid state nanopores to mimic biological
  ion channels: {A} review.
\newblock {\em Adv. Coll. Interf.}, 250:195--213, 2017.

\bibitem{stein_prl_2004}
D.~Stein, M.~Kruithof, and C.~Dekker.
\newblock Surface-charge-governed ion transport in nanofluidic channels.
\newblock {\em Phys. Rev. Lett.}, 93(3):035901, 2004.

\bibitem{Siwy_2004}
Z.~Siwy, E.~Heins, C.~C. Harrell, P.~Kohli, and C.~R. Martin.
\newblock Conical-nanotube ion-current rectifiers:~ the role of surface charge.
\newblock {\em J. Am. Chem. Soc.}, 126(35):10850--10851, 2004.

\bibitem{singh_jap_2011}
K.~P. Singh and M.~Kumar.
\newblock Effect of surface charge density and electro-osmotic flow on ionic
  current in a bipolar nanopore fluidic diode.
\newblock {\em J. Appl. Phys.}, 110(8):084322, 2011.

\bibitem{nasir_acsami_2014}
S.~Nasir, M.~Ali, P.~Ramirez, V.~G{\'{o}}mez, B.~Oschmann, F.~Muench, M.~N.
  Tahir, R.~Zentel, S.~Mafe, and W.~Ensinger.
\newblock Fabrication of single cylindrical {Au}-coated nanopores with
  non-homogeneous fixed charge distribution exhibiting high current
  rectifications.
\newblock {\em {ACS} Appl. Mater. Inter.}, 6(15):12486--12494, 2014.

\bibitem{Zhang_2015}
H.~Zhang, Y.~Tian, J.~Hou, X.~Hou, G.~Hou, R.~Ou, H.~Wang, and L.~Jiang.
\newblock Bioinspired smart gate-location-controllable single nanochannels:
  Experiment and theoretical simulation.
\newblock {\em {ACS} Nano}, 9(12):12264--12273, 2015.

\bibitem{bayley_chem_rev_2000}
H.~Bayley and C.~R. Martin.
\newblock Resistive-pulse sensing from microbes to molecules.
\newblock {\em Chem. Rev.}, 100(7):2575--2594, 2000.

\bibitem{siwy_jacs_2005}
Z.~Siwy, L.~Trofin, P.~Kohli, L.~A. Baker, C.~Trautmann, and C.~R. Martin.
\newblock Protein biosensors based on biofunctionalized conical gold nanotubes.
\newblock {\em J. Am. Chem. Soc.}, 127(14):5000--5001, 2005.

\bibitem{madai_jcp_2017}
E.~M\'adai, M.~Valisk\'o, A.~Dallos, and D.~Boda.
\newblock Simulation of a model nanopore sensor: {Ion} competition underlines
  device behavior.
\newblock {\em J. Chem. Phys.}, 147(24):244702, 2017.

\bibitem{siwy_nim_2003}
Z.~Siwy, P.~Apel, D.~Dobrev, R.~Neumann, R.~Spohr, C.~Trautmann, and K.~Voss.
\newblock Ion transport through asymmetric nanopores prepared by ion track
  etching.
\newblock {\em Nucl. Instrum. Meth. B}, 208:143--148, 2003.

\bibitem{hou_advmat_2010}
X.~Hou, Y.~Liu, H.~Dong, F.~Yang, L.~Li, and L.~Jiang.
\newblock A {pH}-gating ionic transport nanodevice: {Asymmetric} chemical
  modification of single nanochannels.
\newblock {\em Adv. Mater.}, 22(22):2440--2443, 2010.

\bibitem{cervera_ea_2011}
J.~Cervera, P.~Ram{\'i}rez, S.~Mafe, and P.~Stroeve.
\newblock Asymmetric nanopore rectification for ion pumping, electrical power
  generation, and information processing applications.
\newblock {\em Electrochim. Acta}, 56(12):4504--4511, 2011.

\bibitem{zhang_cc_2013}
H.~Zhang, Y.~Tian, and L.~Jiang.
\newblock From symmetric to asymmetric design of bio-inspired smart single
  nanochannels.
\newblock {\em Chem. Commun.}, 49:10048--10063, 2013.

\bibitem{ali_acsami_2015}
M.~Ali, I.~Ahmed, S.~Nasir, P.~Ramirez, C.~M. Niemeyer, S.~Mafe, and
  W.~Ensinger.
\newblock Ionic transport through chemically functionalized hydrogen
  peroxide-sensitive asymmetric nanopores.
\newblock {\em {ACS} Appl. Mater. Inter.}, 7(35):19541--19545, 2015.

\bibitem{zhang_csr_2018}
Z.~Zhang, L.~Wen, and L.~Jiang.
\newblock Bioinspired smart asymmetric nanochannel membranes.
\newblock {\em Chem. Soc. Rev.}, 47(2):322--356, 2018.

\bibitem{Ali_AC_2018}
M.~Ali, I.~Ahmed, P.~Ramirez, S.~Nasir, S.~Mafe, C.~M. Niemeyer, and
  W.~Ensinger.
\newblock Lithium ion recognition with nanofluidic diodes through
  host{\textendash}guest complexation in confined geometries.
\newblock {\em Anal. Chem.}, 90(11):6820--6826, 2018.

\bibitem{Ali_Lang_2017}
M.~Ali, I.~Ahmed, P.~Ramirez, S.~Nasir, J.~Cervera, S.~Mafe, C.~M. Niemeyer,
  and W.~Ensinger.
\newblock Cesium-induced ionic conduction through a single nanofluidic pore
  modified with calixcrown moieties.
\newblock {\em Langmuir}, 33(36):9170--9177, 2017.

\bibitem{Ali_ACSnano_2012}
M.~Ali, S.~Nasir, P.~Ramirez, J.~Cervera, S.~Mafe, and W.~Ensinger.
\newblock Calcium binding and ionic conduction in single conical nanopores with
  polyacid chains: Model and experiments.
\newblock {\em {ACS} Nano}, 6(10):9247--9257, 2012.

\bibitem{liu_JACS_2015}
Q.~Liu, K.~Xiao, L.~Wen, H.~Lu, Y.~Liu, X-Y. Kong, G.~Xie, Z.~Zhang, Z.~Bo, and
  L.~Jiang.
\newblock Engineered ionic gates for ion conduction based on sodium and
  potassium activated nanochannels.
\newblock {\em J. Am.Chem. Soc.}, 137(37):11976--11983, 2015.

\bibitem{ensinger_2018}
W.~Ensinger, M.~Ali, S.~Nasir, I.~Duznovic, C.~Trautmann, M.~E. Toimil-Molares,
  G.~R. Distefano, B.~Laube, M.~Bernhard, M.~Mikosch-Wersching, H.~F. Schlaak,
  and M.~El Khoury.
\newblock The {iNAPO} project: Biomimetic nanopores for a new generation of
  lab-on-chip micro sensors.
\newblock {\em Int. J. Theor. Appl. Nanotech.}, 6:21--28, 2018.

\bibitem{ali_chemcomm_2015}
M.~Ali, S.~Nasir, and W.~Ensinger.
\newblock Bioconjugation-induced ionic current rectification in
  aptamer-modified single cylindrical nanopores.
\newblock {\em Chem. Comm.}, 51(16):3454--3457, 2015.

\bibitem{vlassiouk_jacs_2009}
I.~Vlassiouk, T.~R. Kozel, and Z.~S. Siwy.
\newblock Biosensing with {Nanofluidic} diodes.
\newblock {\em J. Am. Chem. Soc.}, 131(23):8211--8220, 2009.

\bibitem{wu_langmuir_2017}
K.~Wu, K.~Xiao, L.~Chen, R.~Zhou, B.~Niu, Y.~Zhang, and L.~Wen.
\newblock Biomimetic voltage-gated ultrasensitive potassium-activated
  nanofluidic based on a solid-state nanochannel.
\newblock {\em Langmuir}, 33(34):8463--8467, 2017.

\bibitem{nie_chemsci_2015}
G.~Nie, Y.~Sun, F.~Zhang, M.~Song, D.~Tian, L.~Jiang, and H.~Li.
\newblock Fluoride responsive single nanochannel: click fabrication and highly
  selective sensing in aqueous solution.
\newblock {\em Chem. Sci.}, 6(10):5859--5865, 2015.

\bibitem{tian_chemcomm_2010}
Y.~Tian, X.~Hou, L.~Wen, W.~Guo, Y.~Song, H.~Sun, Y.~Wang, L.~Jiang, and
  D.~Zhu.
\newblock A biomimetic zinc activated ion channel.
\newblock {\em Chem. Comm.}, 46(10):1682, 2010.

\bibitem{ali_analchem_2011}
M.~Ali, M.~N. Tahir, Z.~Siwy, R.~Neumann, W.~Tremel, and W.~Ensinger.
\newblock Hydrogen peroxide sensing with horseradish peroxidase-modified
  polymer single conical nanochannels.
\newblock {\em Anal. Chem.}, 83(5):1673--1680, 2011.

\bibitem{hou_materchemA_2014}
G.~Hou, H.~Zhang, G.~Xie, K.~Xiao, L.~Wen, S.~Li, Y.~Tian, and L.~Jiang.
\newblock Ultratrace detection of glucose with enzyme-functionalized single
  nanochannels.
\newblock {\em J. Mater. Chem. A}, 2(45):19131--19135, 2014.

\bibitem{perezmitta_nl_2018}
G.~P{\'{e}}rez-Mitta, A.~S. Peinetti, M.~L. Cortez, M.~E. Toimil-Molares,
  C.~Trautmann, and O.~Azzaroni.
\newblock Highly sensitive biosensing with solid-state nanopores displaying
  enzymatically reconfigurable rectification properties.
\newblock {\em Nano Lett.}, 18(5):3303--3310, 2018.

\bibitem{sun_chemcomm_2012}
Z.~Sun, C.~Han, L.~Wen, D.~Tian, H.~Li, and L.~Jiang.
\newblock {pH} gated glucose responsive biomimetic single nanochannels.
\newblock {\em Chem. Comm.}, 48(27):3282, 2012.

\bibitem{hancsok_2011}
J.~Hancs{\'{o}}k, S.~Kov{\'{a}}cs, Gy. P\"{o}lczmann, and T.~Kasza.
\newblock Investigation the effect of oxygenic compounds on the isomerization
  of bioparaffins over {Pt}/{SAPO}-11.
\newblock {\em Top. Catal.}, 54(16--18):1094--1101, 2011.

\bibitem{berti_jctc_2014}
C.~Berti, S.~Furini, D.~Gillespie, D.~Boda, R.~S. Eisenberg, E.~Sangiorgi, and
  C.~Fiegna.
\newblock A {3-D Brownian Dynamics} simulator for the study of ion permeation
  through membrane pores.
\newblock {\em J. Chem. Theor. Comput.}, 10(8):2911--2926, 2014.

\bibitem{boda_jctc_2012}
D.~Boda and D.~Gillespie.
\newblock Steady state electrodiffusion from the {Nernst-Planck} equation
  coupled to {Local Equilibrium Monte Carlo} simulations.
\newblock {\em J. Chem. Theor. Comput.}, 8(3):824--829, 2012.

\bibitem{hato_jcp_2012}
Z.~Hat\'o, D.~Boda, and T.~Krist\'of.
\newblock Simulation of steady-state diffusion: {Driving} force ensured by
  {Dual Control Volume}s or {Local Equilibrium Monte Carlo}.
\newblock {\em J. Chem. Phys.}, 137(5):054109, 2012.

\bibitem{boda_jml_2014}
D.~Boda, R.~Kov\'acs, D.~Gillespie, and T.~Krist\'of.
\newblock Selective transport through a model calcium channel studied by
  {Local} {Equilibrium} {Monte} {Carlo} simulations coupled to the
  {Nernst}-{Planck} equation.
\newblock {\em J. Mol. Liq.}, 189:100--112, 2014.

\bibitem{boda_arcc_2014}
D.~Boda.
\newblock In R.~A. Wheeler, editor, {\em Ann. Rep. Comp. Chem.}, volume~10,
  chapter 5 {Monte Carlo} Simulation of Electrolyte Solutions in Biology: {In}
  and Out of Equilibrium, pages 127--163. Elsevier, 2014.

\bibitem{hato_cmp_2016}
Z.~Hat\'o, D.~Boda, D.~Gillepie, J.~Vrabec, G.~Rutkai, and T.~Krist\'of.
\newblock Simulation study of a rectifying bipolar ion channel: detailed model
  versus reduced model.
\newblock {\em Cond. Matt. Phys.}, 19(1):13802, 2016.

\bibitem{hato_pccp_2017}
Z.~Hat{\'{o}}, M.~Valisk{\'{o}}, T.~Krist{\'{o}}f, D.~Gillespie, and D.~Boda.
\newblock Multiscale modeling of a rectifying bipolar nanopore: explicit-water
  versus implicit-water simulations.
\newblock {\em Phys. Chem. Chem. Phys.}, 19(27):17816--17826, 2017.

\bibitem{matejczyk_jcp_2017}
B.~Matejczyk, M.~Valisk{\'{o}}, M.-T. Wolfram, J.-F. Pietschmann, and D.~Boda.
\newblock Multiscale modeling of a rectifying bipolar nanopore: {Comparing}
  {Poisson-Nernst-Planck} to {Monte Carlo}.
\newblock {\em J. Chem. Phys.}, 146(12):124125, 2017.

\bibitem{fertig_mp_2018}
D.~Fertig, M.~Valisk\'o, and D.~Boda.
\newblock Controlling ionic current through a nanopore by tuning {pH}: a {Local
  Equilibrium Monte Carlo} study.
\newblock {\em Mol. Phys.}, in press, 2018.

\bibitem{madai_pccp_2018}
E.~M\'adai, B.~Matejczyk, A.~Dallos, M.~Valisk\'o, , and D.~Boda.
\newblock Controlling ion transport through nanopores: modeling transistor
  behavior.
\newblock {\em Phys. Chem. Chem. Phys.}, 20(37):24156--24167, 2018.

\bibitem{metropolis}
N.~Metropolis, A.~W. Rosenbluth, M.~N. Rosenbluth, A.~H. Teller, and E.~Teller.
\newblock Equations of state calculations by fast computing machines.
\newblock {\em J. Chem. Phys.}, 21(6):1087--1092, 1953.

\end{thebibliography}
% \bibliographystyle{unsrt} %the RSC's .bst file

\end{document}